\documentclass[pra,twocolumn,a4paper,showpacs,superscriptaddress]{revtex4}
\usepackage{amsmath}
\usepackage{amsfonts}
\usepackage{graphicx}
\usepackage{natbib}
\usepackage{exscale}
\usepackage{amssymb}
\usepackage{bbold}
\newcommand{\ket}[1]{\ensuremath{|\,#1\,\rangle}}
\newcommand{\bra}[1]{\ensuremath{\langle\,#1\,|}}

\newcommand{\mf}[1]{\mbox{\boldmath$ #1 $\unboldmath}}

\begin{document}

\title{Detection of atomic entanglement and electromagnetically
   induced transparency in velocity-selective coherent population 
trapping}
\author{M. Kiffner}
\affiliation{Fachbereich Physik der Universit\"at Konstanz, 
Fach M674, 78457 Konstanz, Germany}
\altaffiliation{Present address: Max-Planck-Institut f\"ur Kernphysik, 
Saupfercheckweg 1, 69117 Heidelberg, Germany}
\author{K.-P. Marzlin}
\affiliation{Department of Physics and Astronomy,
University of Calgary,
2500 University Drive NW, Calgary, Alberta T2N 1N4, Canada}
\affiliation{Fachbereich Physik der Universit\"at Konstanz, 
Fach M674, 78457 Konstanz, Germany}


\begin{abstract}
We investigate theoretically the optical properties of an atomic gas which 
has been cooled by the laser cooling method velocity-selective coherent 
population trapping. We demonstrate that the application
of a weak laser pulse gives rise to a backscattered pulse, which is a direct 
signal for the entanglement in the atomic system, and which leads to
single-particle entanglement on the few-photon level. If the pulse is
applied together with the pump lasers, it also displays the phenomenon
of electromagnetically induced transparency. We suggest that the effect
should be observable in a gas of Rubidium atoms.
\end{abstract}

\maketitle

\section{Introduction}
Among a large variety of laser cooling schemes that have been developed, 
velocity-selective coherent population trapping (VSCPT) belongs to 
a small group of methods by which temperatures below the one-photon 
recoil energy can be achieved. One-dimensional realizations of the 
VSCPT-method  have been demonstrated  for ${}^{4}$He \cite{vscptexp1} 
and  ${}^{87}$Rb atoms \cite{haensch}. In addition, VSCPT experiments 
with Helium atoms have been carried out successfully  in 
two \cite{vscptexp2} and three \cite{vscptexp3} dimensions. 

The fundamental feature on which VSCPT relies is the preparation
of an atomic dark state (see, e.g., Ref.~\cite{stirap})
which does not couple to the pump lasers.
However, there is a significant difference between conventional
dark states and the state of a VSCPT gas:
In the former case it is preferable 
to consider situations in which the atomic center-of-mass motion 
can be neglected. This is for example the case if all laser fields 
are propagating in one direction,  
the corresponding dark state is then simply a superposition of
two hyperfine ground states.  By contrast, the dark state of a 
VSCPT gas must depend on the center-of-mass motion to achieve 
the desired cooling effect. Therefore, the atoms are exposed to 
two counter-propagating pump lasers with wavenumber $k_p$.
The  dark state of this laser configuration 
\begin{equation} 
\ket{\text{VSCPT}} := \frac{1}{\sqrt{2}}\big[\,
  \ket{-}\otimes\ket{-\hbar k_p} - \ket{+}\otimes\ket{\hbar k_p}\,\big]
\end{equation} 
is then an entangled superposition of states 
that contain both  internal ($\ket{\pm}$) and center-of-mass 
degrees of freedom ($\ket{\pm \hbar k_p}$). 

It is this existence of entangled atomic coherences which makes the 
optical properties of a VSCPT gas interesting. 
Our goal is to identify a distinctive optical signal which is directly
linked to the atomic entanglement. We will show that the latter gives 
rise to a backscattered beam of light 
when the VSCPT gas is probed with a weak signal laser pulse 
(Sec.~\ref{backscatter}). 
It will be demonstrated that this signal is absent for a mere
mixture of ground states and thus would provide a direct test of
the entanglement of the atomic system. 
This effect can also be interpreted
as a transfer of entanglement from a single atom to a single photon
(Sec.~\ref{secqft}). There are two additional features of the VSCPT state
which make it a system with very special optical properties. First,
since it is prepared in a dark state one can expect that a VSCPT gas 
also exhibits the phenomenon of electromagnetically induced transparency 
(EIT) \cite{hau1,fleischhauer1,phillips1}; this  
will be examined in  Sec.~\ref{eit}. 
We remark here that EIT in a standing-wave geometry has also been 
studied by Affolderbach {\em et al.} \cite{affolderbach}. 
Although  the laser configuration is similar to that considered here,
the physical system is quite different since 
the experiments are conducted at room temperature. 
A second feature of the VSCPT state is that it is a periodic state of matter
in position space 
($\langle z \ket{\text{VSCPT}} \propto \exp (\pm i p z/\hbar)$).
While one may conjecture that this would cause the creation of
band gaps in the photonic spectrum it will become clear in our derivations
that this is not the case.

\section{Theory of the VSCPT state \label{system}}
\begin{figure}[b!]
\begin{center}
\includegraphics{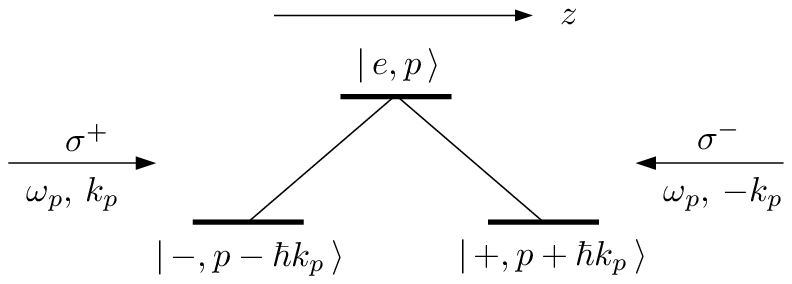}
\caption{\label{lambda}
$\Lambda$-type pumping scheme including translational degrees of freedom. 
A $\sigma^+$ polarized beam travels along the 
positive $z$-direction  and 
couples the ground state $|-,p-\hbar k_p \rangle$ to 
the excited state $|e,p \rangle$. 
The counter-propagating beam  
is  $\sigma^-$ polarized  and couples 
the ground state $|+,p+\hbar k_p\rangle $ to the excited state.}
\end{center}
\end{figure}

Our aim is to find an optical signature for the entanglement 
between the atomic internal and center-of-mass degrees of freedom
in a VSCPT state, as well as to explore its potential for EIT
effects. 
To do so we first describe the features of an atomic VSCPT state,
which has been experimentally realized with 
\mbox{${}^4$He} \cite{vscptexp1} as well as \mbox{${}^{87}$Rb} 
atoms \cite{haensch}.
The atoms are exposed to
two counter-propagating pump laser beams which share the
same frequency $\omega_p$ 
and Rabi frequency $\Omega_p$. 
The electronic degrees of freedom of the atoms  
are modeled by a three-level system in $\Lambda$-configuration
(see Fig.~\ref{lambda}). In addition, the atomic center-of-mass motion is 
treated quantum mechanically. In the one-dimensional 
cooling scheme considered here the momentum components perpendicular to 
the pump field are not observed and can therefore be traced out. 
The atomic Hamiltonian is then given by
\begin{equation}
H_0 := \frac{\hat{p}_z^2}{2\,m}\,  +\,\hbar\,\omega_0\,\ket{e}\bra{e} \; ,
\end{equation}
where $m$ is the mass of the atom and $\omega_0$ its resonance frequency.
In the rotating wave approximation the interaction Hamiltonian
takes the form
\begin{eqnarray}
H_p     
& := &    -\,\left(\, \ket{e}\bra{-} 
\,e^{i k_p \hat{z}}\hspace*{1.2cm} \right. \label{hp} \\[0.2cm] 
& & \qquad\quad \left. + \, \ket{e} \bra{+}\,e^{-i k_p \hat{z}}\,\right)\,
e^{-i\omega_p t}\,\hbar\,\Omega_p \,+\, \text{H.c.} \; ,
\nonumber
\end{eqnarray}
where $\Omega_p := d E_p/\hbar$ is the Rabi frequency of the pump lasers
($d := \bra{e}\mf{\hat{d}}\cdot\mf{\epsilon}^{(\pm)}\ket{\mp}$
and $\mf{\hat{d}}$ is the electric-dipole moment operator). 
$E_p$ denotes the electric-field amplitude of the classical pump field 
$\mf{E}_p(z,t) = \mf{E}_p^{(+)}(z,t) +  \mf{E}_p^{(-)}(z,t)$, with
\begin{eqnarray}
\mf{E}_p^{(+)}(z,t) & := & E_p e^{-i \omega_p t} \left (
  \mf{\epsilon}^{(+)}
  e^{i k_p z }  
  + \mf{\epsilon}^{(-)} e^{-i k_p z } \right )
  \label{epump}
\end{eqnarray}
being the positive-frequency part of the electric field and 
$\mf{E}_p^{(-)} =  \mf{E}_p^{(+)\dagger}$. The circular polarization vectors
are defined as $\mf{\epsilon}^{(\pm)}:=\left(\mf{e}_x 
\pm i\mf{e}_y\right)/\sqrt{2}$\,.

The special linear combination of ground states 
\begin{equation}
\ket{D(p)} := \frac{1}{\sqrt{2}}\,\big[\,\ket{-,p-\hbar k_p} \,-\,
\ket{+,p + \hbar k_p}\,\big]
\label{ds}
\end{equation}
is an eigenstate of the interaction Hamiltonian $H_p$ and therefore 
decoupled from the light field. 
However, $\ket{D(p)}$ is generally not an eigenstate of 
$H_0$. Only for $p=0$ the state
$
\ket{\mbox{\small VSCPT}}   =  \ket{D(p=0)} \label{stateideal}
$
is an eigenstate of the complete Hamiltonian 
$
H_{\text{C}}(t) :=  H_0 + H_p(t) 
$
and therefore stationary.

In principle, a VSCPT state can simply be created by pumping the 
atoms with laser
light. Since spontaneous emission increases the population in
$|\text{VSCPT}\rangle$ with a certain probability, 
and since all other combinations
of the ground states can be pumped back to the excited state, the atoms
accumulate in $|\text{VSCPT}\rangle$. 
The corresponding dynamics
is governed by the master equation 
\begin{equation}
  \dot{\varrho} = - \frac{i}{\hbar}\,[H_{\text{C}}(t),\varrho] 
  +\mathcal{L}_{\gamma}\varrho \quad.
\label{masterequation}
\end{equation}
The term $\mathcal{L}_{\gamma}\varrho$ on the right hand side accounts 
for spontaneous emission and is given by \cite{vscpttheorie,spontan,stenholm}
\begin{eqnarray}
  \mathcal{L}_{\gamma}\varrho & := & -\gamma\varrho_{ee} \ket{e}\bra{e}
   \label{gammaone}\\[0.2cm]
  & & -\frac{\gamma}{2}\,\left[\varrho_{e-}\ket{e}\bra{-} \,+ \,
   \varrho_{e+}\ket{e}\bra{+} \, + \, \text{H.c.}\right] \nonumber \\
  & & \hspace{0.7cm} +\, \frac{\gamma}{2}   \int
  \limits_{-\hbar k_0}^{\hbar k_0} H(u)\,
   e^{\frac{i}{\hbar}u \hat{z}}\varrho_{ee} e^{-\frac{i}{\hbar}
u \hat{z}}\,\text{d}u\,\ket{-}\bra{-} 
\nonumber \\[0.2cm]
& & \hspace{0.7cm} + \, \frac{\gamma}{2}   \int\limits_{-\hbar k_0}^{\hbar k_0} H(u)\,
e^{\frac{i}{\hbar} u \hat{z}}\varrho_{ee} e^{-\frac{i}{\hbar} u \hat{z}}
  \,\text{d}u\, \ket{+}\bra{+} \nonumber
\end{eqnarray}
with 
$ \varrho_{\epsilon\epsilon'} := 
  \bra{\epsilon}\varrho\ket{\epsilon'}\quad (\epsilon,
  \,\epsilon'\,\in\,\{e,\,-,\,+\})$.
For the $J_e=1\leftrightarrow J_g=1$ transition considered here, the 
function $H(u)$ is given by $H(u) := 3\,
\,(1 + u^2/(\hbar k_0)^2)/(8\,\hbar k_0)$.
It has been shown in 
Ref.~\cite{vscpttheorie} that for a finite duration of the pumping 
this process results in a finite atomic momentum distribution.
We approximate this state by a mixture of dark states with different momenta
that is described by the density matrix \cite{remark1} 
\begin{equation}
  \varrho_f  :=  \int\limits_{-\infty}^{\infty}  
  f(p)\,\ket{D(p)}\bra{D(p)}\,\text{d}p \; ,
\label{rhop}
\end{equation}
where $f$ characterizes the momentum distribution.
We assume here that $f$ can be approximated by a Gaussian distribution 
centered around $p=0$,
\begin{equation}
f(p) = \frac{1}{\sqrt{2 \pi}\, \sigma_p}\,e^{- p^2/(2\sigma_p^2)}\; .
\label{deff}
\end{equation}
The momentum width is given by $\sigma_p^2 = \int p^2 f(p) \text{d}p $. It
is a measure of the achieved final 
temperature of the gas and varies with the coherent interaction time $\Theta$ 
as  $\sigma_p \sim 1/\sqrt{\Theta}$. The experiments presented in 
Refs.~\cite{vscptexp1,haensch} demonstrate that a value of
$\sigma_p \approx \hbar k_p/2$ is achievable.

\section{Coherent backscattering of a weak signal beam \label{backscatter}}
The optical response of an atomic gas can be described
by the Maxwell-Bloch equations, which include an atomic master
equation of the form (\ref{masterequation}) and the wave equation
for the electric field, 
\begin{equation}
   \left(\frac{1}{c^2}\,\partial_t^2\,-\,\Delta\right)\mf{E} = 
   -\frac{1}{c^2\,\varepsilon_0}\, \ddot{\mf{P}} \; .
\label{wellengl}
\end{equation}
For a dilute gas, atom-atom interactions 
can be neglected so that the macroscopic polarization $\mf{P}$ 
can be expressed as the local mean value of the single-atom dipole 
operator,
\begin{equation}
\mf{P}(z,t):= \bar{\rho}  \,
  \mbox{Tr}_{\text{int}}\big(\varrho(z,t) \mf{\hat{d}}\big) \; .
\label{meanPolarization}
\end{equation}
Here, $\bar{\rho}$ denotes the mean atomic 
density and the trace runs over the internal states 
($\varrho(z,t) = \bra{z}\varrho(t)\ket{z}$).  
\begin{figure}[t!]
\begin{center}
\includegraphics{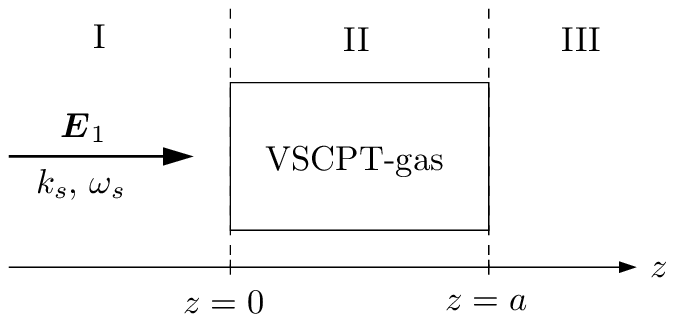}
\caption{\label{rand} Idealized set-up after the pump laser beams have 
been switched off. The VSCPT gas  stretches from $z=0$ to 
$z=a$ (region II) and is surrounded by vacuum (region I and III). 
The signal beam, a plane-wave with frequency $\omega_s$, $\sigma^+$ 
polarization and amplitude $E_0$ is shone into the probe from the left.}
\end{center}
\end{figure}

To investigate the optical properties of a gas that has been 
cooled by means of the VSCPT method, we generally consider
the behavior of a weak signal laser beam interacting with a
homogeneous distribution of cooled atoms of finite width $a$.
In this section we examine the case where the signal beam
is switched on after the pump lasers are switched off. We will see in Sec.~\ref{eit} 
that the optical response will be quite different
when the pump lasers are not switched off.

The setup to investigate the optical response
is shown in Fig.~\ref{rand}. 
At time $t=0$, the pump field is turned off and a $\sigma^+$ polarized 
signal field 
\begin{equation}
  \mf{E}_1^{(+)}(z,t) = E_1(z) \mf{\epsilon}^{(+)}
   e^{i (k_s z - \omega_s t)}
  \label{ansatzefeld}
\end{equation}
with slowly varying amplitude $E_1(z)$ and frequency $\omega_s$  
is applied to the probe from the left. 
The associated atomic evolution can be derived from Eq.~(\ref{masterequation})
with $H_\text{C}$ replaced by $H_0 + H_\text{s}(t)$, with 
\begin{eqnarray}
H_\text{s}(t) &=&  
  - \hbar \Omega_1\,e^{i k_s \hat{z}}\,e^{ - i \omega_s t} 
   |e \rangle\langle -|
  \nonumber \\ & &
  - \hbar \Omega_2\,e^{i k_2 \hat{z}}\,e^{ - i \omega_s t} 
   |e \rangle\langle +|
   \; + \; \text{H.c.}
\label{hs}
\end{eqnarray}
and $\Omega_i := d\, E_i(z)/\hbar$. The second electric field
\begin{equation}
  \mf{E}_2^{(+)}(z,t) := E_2(z)\,\mf{\epsilon}^{(-)}
  e^{i ( k_2 z - \omega_s t)}   
\label{ansatze2}
\end{equation}
($k_2 := k_s -2 k_p\approx -k_s $)
has been introduced to keep the ansatz consistent (see below).
To solve the Maxwell-Bloch equations we first consider the 
atomic master equation. Since the signal beam is assumed to be weak,
its influence on the atoms can be treated in first-order perturbation 
theory. Expanding the density matrix as $\varrho = \varrho_0  + \varrho_1 $
we then find 
\begin{eqnarray}
  \dot{\varrho}_0 &=& \mathcal{L}_0\varrho_0 \label{zeroorder} \\
  \big[\partial_t \,-\,\mathcal{L}_0\big]\varrho_1  & = & 
  - \frac{i}{\hbar}\left[H_\text{s},\varrho_0\right] \;.
\label{firstorder} 
\end{eqnarray}
The Liouville operator $\mathcal{L}_0$ is defined as 
\begin{equation}
\mathcal{L}_0(\,\cdot\,):=-\frac{i}{\hbar}\,[H_0,\,\cdot\,] \, + \, 
\mathcal{L}_{\gamma}(\,\cdot\,) 
\label{L0def}
\end{equation}  
and governs the time evolution of the free atom.

We first consider the ideal case where $\varrho_0$ is given by
the stationary state 
$\varrho_{\text{s}} \equiv|\text{VSCPT} \rangle \langle \text{VSCPT}|$.
For brevity we will only discuss the incoming signal beam $\mf{E}_1$
since the second beam can be treated in an analogous way
\cite{linearity}.  
The commutator on the right hand side of eq. (\ref{firstorder}) 
is comprised of two time-dependent parts 
\begin{equation} 
  -\, \frac{i}{\hbar}\left[H_\text{s},\varrho_{\text{s}}\right] = 
   \frac{i\Omega_1}{\sqrt{2}}
   |e,\hbar\Delta k \rangle \langle \text{VSCPT}|
  e^{-i\omega_s t}\, + \, \text{H.c.}
\end{equation}
that vary with $e^{-i\omega_s t}$ and $e^{i\omega_s t}$, respectively 
($\Delta k := k_s - k_p $).
This inhomogeneity gives rise to the following steady-state solution 
\begin{eqnarray}
  \varrho_1 (t) & = &  -\,\frac{1}{2}\,\chi_0(\omega_s)\,
  \Omega_1\,e^{-i\omega_s t}\,\ket{e,\hbar\Delta k}\bra{-, -\hbar k_p}  
  \label{rhoe1} \\[0.2cm]
  & & +  \,\frac{1}{2}\,\chi_0(\omega_s)\,\Omega_1\,e^{-i\omega_s t}\,
  \ket{e,\hbar\Delta k}\bra{+ , \hbar k_p} \, + \,\mbox{H.c.}\nonumber
\end{eqnarray}
with
\begin{equation}
  \chi_0(\omega_s)  :=  \left[\, \left( E_r - 
  \frac{\hbar {\Delta k}^2}{2 m}\right) 
  + i\,\frac{\gamma}{2}  + \Delta_s\,\right]^{-1} \quad.
\label{susi} 
\end{equation}
$E_r := \hbar k_p^2 /(2 m)$ defines the recoil frequency and
$\Delta_s := \omega_s - \omega_0$ denotes 
the detuning of the signal field from  resonance.
In the following we will use this steady-state solution instead of
a full solution that fulfills the correct initial condition $\varrho_1(t)=0$. 
This is justified if the interaction
time between the laser (pulses) and the atoms is long compared
to the natural lifetime of the excited state, since in this case
all non-stationary contributions (which are solutions of the homogeneous
equations) are damped away.

We remark that, as a consequence of first-order perturbation theory,
the population of the excited state remains zero. This explains the
simple form of solution (\ref{rhoe1}) and effectively allows us to
replace the full decoherence term (\ref{gammaone}) by 
\[
\mathcal{L}_{\gamma}^{\text{\small coh}}\varrho := 
 -\frac{\gamma}{2}\,\left[\varrho_{e-}\ket{e}\bra{-} +
\varrho_{e+}\ket{e}\bra{+} + \mbox{H.c.}\right] \; .
\]

An important feature of Eq.~(\ref{rhoe1}) is the non-vanishing
coherence $[\varrho_1]_{e+}:=\bra{e}\varrho_1\ket{+}$.
This effect does not appear if the state 
$\varrho_0=|\text{VSCPT}\rangle\langle\text{VSCPT}|$ is replaced by
an incoherent mixture of the form
\[
  \varrho_{\text{mix}} := \frac{1}{2}
  \big[ \ket{-,-\hbar k_p}\bra{-,-\hbar k_p} 
   + \ket{+,\hbar k_p}\bra{+,\hbar k_p}  \big] \; ,
\]
and therefore is a signal for the coherence between the atomic
ground states. 

We will now show that this coherence 
creates a backscattered light beam and therefore  generates
a signal of the entanglement between atomic internal and center-of-mass
degrees of freedom. To do this we solve the wave equation
(\ref{wellengl}) in  paraxial approximation, i.e., we neglect
the terms $\partial_z^2 E_i(z)$ ($i \in \{1,2\}$). This is justified if
the envelopes $E_i(z)$ are slowly varying over one wavelength, so that
\begin{equation}
  \left|\partial_z^2 E_i(z)\right|  \ll  
  k_s\,\left| \partial_z E_i(z)\right| \; .
\label{paraxial}
\end{equation}

Inserting $\varrho_1$ of Eq.~(\ref{rhoe1}) and the 
corresponding contribution induced by $\mf{E}_2$
into Eq.~(\ref{meanPolarization}) leads to
\begin{eqnarray}
  \mf{P}^{(+)}(z,t) 
  &=& \frac{ \bar{\rho} d^*}{2 } \chi_0(\omega_s)  
  \left(  \big( \Omega_2 - \Omega_1 \big)  e^{i k_s z } 
  \mf{\epsilon}^{(+)} \right.  
  \label{fullp} \\ & & \hspace{0.8cm} 
  \left. +  \big( \Omega_1 - \Omega_2 \big) e^{i k_2 z }
  \mf{\epsilon}^{(-)}\right) e^{-i\omega_s t}   \; .
\nonumber 
\end{eqnarray}
The term $\sim \Omega_1 \mf{\epsilon}^{(-)}$
is a direct consequence of $[\varrho_1]_{e+}\neq 0$. 
In the paraxial wave equation \cite{neglect}, 
\begin{eqnarray}
  k_s  \partial_z E_1(z) \mf{\epsilon}^{(+)} e^{i  k_s z} \hspace{2.4cm} & & \\
  -\,k_s\big( \partial_z E_2(z)+2i\Delta k E_2(z)\big)\mf{\epsilon}^{(-)} e^{i  k_2 z}
  & = &  
   \frac{e^{i\omega_s t}}{2 i c^2\,\varepsilon_0}\, \ddot{\mf{P}}^{(+)} \; ,\nonumber
\end{eqnarray}
this results in a coupling between $E_1$ and $E_2$. If $E_2$
had not been introduced, the term  $\sim \Omega_1 \mf{\epsilon}^{(-)}$
would not have a corresponding term  $\sim\mf{\epsilon}^{(-)}$ on the
left-hand side, so that the equation would be inconsistent. Sorting
the terms according to their polarization and phase factors leads to
a coupled equation for the amplitudes,
\begin{equation}
  \partial_z \left(\begin{array}{c}
    E_1 \\ E_2 \end{array} \right) = 
  \left(\begin{array}{cc}
    -i n_0 & i n_0 \\
    -i n_0 & i(n_0  - 2\,\Delta k)
  \end{array}\right)
  \left(\begin{array}{c}
    E_1 \\ E_2 \end{array} \right) \; ,
\label{dglsystem}
\end{equation}
where
\begin{equation}
n_0(\omega_s)   :=    \frac{1}{2\,k_s}  \,
\frac{\omega_s^2 \,\bar{\rho} \,|d|^2}{2 c^2 \varepsilon_0 \hbar}\,  
\chi_0(\omega_s) \quad. \label{nnull}
\end{equation}
\begin{figure}[b!]
\begin{center}
\includegraphics[scale=0.65]{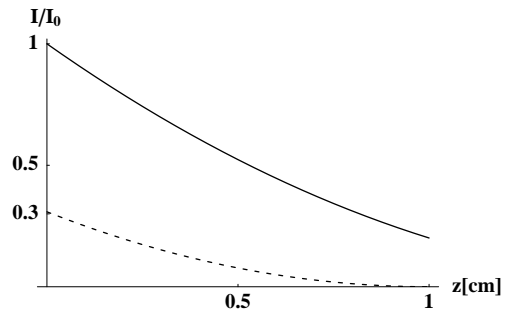}
\caption{\label{intensity} \small Intensity of the incoming 
(solid line) and the reflected (dashed line) beam in a VSCPT-gas 
of length $a=1\,$cm. Both intensities are related to the intensity 
$I_0$ of the incoming beam in region I (Fig. \ref{rand}).}
\end{center}
\end{figure}

The boundary conditions  on the physical solution are determined
by the behavior of $E_1$ and $E_2$ outside the gas and
Maxwell's equation 
$\nabla\times\mf{E} = - \dot{\mf{B}}$, which
implies that the transverse electric field 
is continous at the boundary of the gas (see Fig.~\ref{rand}). 
Since $E_1$ corresponds to the incoming signal beam that travels
along the positive $z$-axis, its amplitude is given by 
a fixed value $ E_1(z) = E_0$ for $z<0$ (region I). On the other hand,
the counter-propagating field $E_2$ is initially empty so that
$E_2(z)=0$ for $z>a$ (region III). 
The boundary conditions for the amplitudes $E_{1,2}$ in region II are thus 
given by \cite{boundary}
\begin{equation}
  E_1(z = 0) \stackrel{!}{=} E_0 \quad\text{and}\quad  
  E_2(z=a)\stackrel{!}{=} 0 \; .
\label{boundary}
\end{equation}
Introducing the notation $\delta := \sqrt{(2 n_0 - \Delta k)\Delta k}$ 
the full solution of Eq.~(\ref{dglsystem}) is found to be
\begin{eqnarray}
    E_1(z) & = &  \label{periodisch} \\
  & & \hspace*{-1.5cm} E_0 \,\frac{\delta\cosh[\delta(a-z)] + 
    i(n_0 -\Delta k)\sinh[\delta(a-z)]}
    {\delta \cosh[\delta a] + i (n_0 - \Delta k) \sinh[\delta a]}
    e^{-i \Delta k z} 
  \nonumber \\[0.2cm]
    E_2(z) & = & E_0\, \frac{i\, n_0 \sinh[\delta(a-z)]}
    {\delta \cosh[\delta a] + i (n_0 - \Delta k) \sinh[\delta a]}
    e^{-i \Delta k z}  
   \; . \nonumber
\end{eqnarray}
The envelopes $E_{1,2}$ contain a periodic part whose period $\lambda$ is 
given by the inverse imaginary part of 
$\delta$. For atomic densities lower than $10^{12} \text{cm}^{-3}$
the parameter $\lambda$ is usually
orders of magnitudes larger than the typical length 
of a VSCPT gas (i.e., a few centimeters); the periodicity would
then not be observable. Generally, for $|a \delta| \ll 1$
and $\Delta k \rightarrow 0$
Eq. (\ref{periodisch}) simplifies considerably and becomes
\begin{eqnarray}
E_1(z) & = & E_0\,\left(1 - \frac{  n_0 z}{ n_0  a -i }\right) \label{linear} \\[0.2cm]
E_2(z) & = & E_0\,\left(\frac{ n_0(a-z)}{ n_0 a - i }\right)\; ,
\nonumber
\end{eqnarray}
which is an excellent approximation for realistic parameters.

Fig. \ref{intensity} shows a plot of the intensity of the 
incoming and reflected beam. 
We chose a probe length of $a=1\,$cm and a detuning of 
$\Delta_s = 3\times 10^6\,s^{-1}$. For the 
$J_e=1\, \leftrightarrow\,J_g=1$ transition of the $D_1$ Line of ${}^{87}$Rb, 
the recoil frequency, the rate of spontaneous emission and the dipole moment take  
the values \cite{rb}
$E_r = 2.3\times10^{4}\,s^{-1}$, $\gamma = 3.61\times 10^7\,s^{-1}$ and $d=2.99\,
e \, a_0  /\sqrt{12}$, 
where $e$ is the elementary charge and $a_0$ is Bohr's radius. 
In addition, we assumed a mean atomic density 
$\bar{\rho}$ of $2\times 10^{10}\,\text{cm}^{-3}$ which is close to the 
experimental conditions  ($\bar{\rho}\approx 10^{10}\,\text{cm}^{-3}$) 
described in Ref.~\cite{haensch}. 
About 30\% of the incoming intensity is transferred to the 
reflected beam, all losses are due to spontaneous emission. 

{\em Simplified explanation of the backscattered beam:}
Since the above derivation is somewhat involved we will give here
another, more physical explanation of the effect. The incoming 
signal beam transfers the initial
atomic state $\ket{\text{VSCPT}}$ to the superposition
$|\psi_1 \rangle \sim \ket{\text{VSCPT}} +  H_s \ket{\text{VSCPT}}
 \sim \ket{\text{VSCPT}} + |e,p=0 \rangle $
(for simplicity we here set $\Delta k=k_s-k_p$ to zero). 
Hence, the absorption of an incoming signal photon
leads to a (partial) transfer of the initial coherence
between $| -, -\hbar k_p \rangle $ and $| +, \hbar k_p \rangle $ 
to one between $|e,0 \rangle$ and 
$|+, \hbar k_p \rangle $. The latter corresponds to an induced dipole moment
which, due to angular momentum conservation, can only lead to emission of
photons with polarisation  $\sigma^-$. 
In Maxwell's equations such photons are coupled to the 
coherence $\rho_{e+}(z,z) = 
\langle e, z| \psi_1 \rangle \langle \psi_1 | + , z \rangle$  
at the position z of the atom. An elementary calculation leads to
$\rho_{e+}(z,z) $
$\sim \langle e, z \nobreak{| e,0 \rangle} \langle +, \hbar k_p| + , z\rangle
\sim \exp (-i k_p z)$ so that the emitted
photons are propagating in the opposite
direction of the incoming beam. 
The associated change in the photon's momentum
is provided by the different atomic momenta in the two ground states.
The $\sigma^-$ polarized backscattered light 
beam $\mf{E}_2$ is thus not only a signal for the 
coherence of the VSCPT state, but also a 
direct signal of the entanglement between atomic internal and center-of-mass
degrees of freedom. 

This method of probing entanglement in atomic gases should not only be
applicable to a VSCPT gas but to any state in which an entanglement
between internal states and momentum states is achieved. 
The VSCPT gas is only special in that the entangled state is also a dark state
with respect to the pump laser field. We will see below that this
will lead to EIT for the pair of signal fields.  
We also emphasize that 
a mere mixture
of states with different momentum and internal degrees of freedom,
as it is created by other cooling methods such as Raman cooling
\cite{raman92}, would not display the backscattered beam. 

\section{Finite momentum width: dephasing of the reflected beam
\label{dephasing}}
In reality the atoms are not prepared in the ideal state
$| \text{VSCPT} \rangle \langle \text{VSCPT} |$ but have a finite
momentum width, described by 
$\varrho_f$ of Eq.~(\ref{rhop}). Since this is not a stationary state,
the free time evolution of the unperturbed density operator 
$\varrho_0$ of Eq.~(\ref{zeroorder}) 
after the pump lasers are switched off 
reads
\begin{equation} 
\varrho_0(t)  \,= 
  \int_{-\infty}^{\infty}  f(p)\,  \ket{D(p,t)}  \bra{D(p,t)}  
\, \text{d}p \; ,
\end{equation} 
where
\begin{equation} 
  \ket{D(p,t)} = \frac{| -,p-\hbar k_p \rangle }{\sqrt{2}}
  - e^{-i\omega_r(p) t} \frac{| +,p+\hbar k_p \rangle }{\sqrt{2}}
\label{darkStateEvolve}\end{equation} 
and $\omega_r(p) := 2k_p p/m$. The energy 
$\hbar \omega_r(p)$ is just the 
difference between the kinetic energies of the 
ground states $\ket{-,p-\hbar k_p}$  and  
$\ket{+,p+\hbar k_p}$.

$\varrho_0(t)$ describes a mixture of initially dark states
which evolve into a corresponding bright state at rates which
depend linearly on $p$. To demonstrate that this behavior 
results in a dephasing of the backscattered wave we
proceed as in section \ref{backscatter} to derive the macroscopic 
polarization (we only write down the part induced by $\mf{E}_1$) 
\begin{equation} 
 \mf{P}^{(+)}(z,t)  =  -\frac{ \bar{\rho} d^* }{2}    \Omega_1
   \Big(   I_{\alpha}\, e^{i k_s z } \,\mf{\epsilon}^{(+)}
   -I_{\beta}  e^{-i k_s z }
   \mf{\epsilon}^{(-)}\Big) e^{-i\omega_s t} 
\label{polarisation2} \end{equation} 
with
\begin{eqnarray} 
  I_{\alpha}  &:=&  \int\limits_{-\infty}^{\infty}\chi(\omega_s,p) f(p) 
  \text{d}p \; , 
  \nonumber \\[0.2cm]
  I_{\beta}  &:=&    \int\limits_{-\infty}^{\infty} \chi(\omega_s,p)
  f(p) e^{i \omega_r(p) t}\text{d}p   
  \nonumber \\[0.2cm]
  \chi(\omega_s,p) & := &  \left[
  E_r  + i\frac{\gamma}{2}  + \Delta_s - \frac{1}{2}
  \omega_r(p)\right]^{-1} \; .
\nonumber
\end{eqnarray}
In the derivation we have used the approximations
$E_r - \hbar \Delta k^2/(2 m) \approx E_r$ and
$k_p \pm \Delta k  \approx k_p$ as well as $k_2\approx -k_s$.
The evaluation of the  integrals  $I_{\alpha}$ and $I_{\beta}$ can be
simplified by using that $\chi(\omega_s,p)$ as a function of
$p$ is almost constant over the momentum range of a VSCPT gas. 
We assume here that the momentum width is given by $\sigma_p = \hbar k_p/2$ 
(see Sec.~\ref{system}) and 
are thus allowed to replace $\chi(\omega_s,p)$ by $\chi_0(\omega_s)$, 
which yields
\begin{equation} 
I_{\alpha}   \approx   \chi_0(\omega_s) 
\quad , \quad
I_{\beta}   \approx   \chi_0(\omega_s) \, g \;,
\end{equation}
with $g(t) := e^{-2\,E_r^2\,t^2}$. 

These results enable us  to describe 
the long-term behavior of the system.  For $t \gg 1/E_r$, the 
Integral $I_{\beta}$ is approximately zero and   the term proportional to 
$\mf{\epsilon}^{(-)}$ in eq. (\ref{polarisation2}) can be neglected. 
Consequently, the backscattered wave is equal to zero. 
The reason is that the dephasing of the coherences in
different dark states $|D(p,t)\rangle $ of (\ref{darkStateEvolve}) leads to a
destructive interference between the backscattered signal beams
for different momenta.

Although the backscattered signal is suppressed for continuously
operating signal beams, it is reasonable to expect that it
will be strong enough if one employs signal pulses instead.
Incoming and backscattered pulses can formally be described by 
assuming that the amplitudes $E_{1,2}$ of
Eqs.~(\ref{ansatzefeld}) and 
(\ref{ansatze2}) are slowly varying both in space and in time, 
\begin{eqnarray}
\left|\partial_z^2 E_{1,2}(z,t)\right|  & \ll &  
k_s\,\left| \partial_z E_{1,2}(z,t)\right| \;  ,
 \nonumber \\[0.2cm]
\left|\partial_t^2      E_{1,2}(z,t)\right| &  \ll &   
\omega_s\, \left|\partial_t E_{1,2}(z,t)\right| 
\ll \omega_s^2\,\left|E_{1,2}(z,t)\right|\; .
\nonumber
\end{eqnarray}
Maxwell's equation (\ref{wellengl}) becomes now a 
coupled set of first-order partial differential 
equations \cite{vgroup} for the envelopes $E_1$ and $E_2$,
\begin{eqnarray}
\left(\frac{1}{c}\,\partial_t\, +  \,\partial_z \right) E_1  & = &  
i n_0 \,\big( E_2 \,g  - E_1  \big) 
\label{partdgl} \\[0.2cm]
\left(\frac{1}{c}\,\partial_t\, -  \,\partial_z \right) E_2  & = &
i n_0 \, \big( E_1\,g - E_2  \big) \quad,
\nonumber
\end{eqnarray}
where $n_0$ is defined in Eq.~(\ref{nnull}). 

We have numerically solved these equations for a Gaussian
amplitude $E_1$ incoming from the left, 
with the boundary condition that $E_2$ is zero on the right-hand side
of the medium, and that the fields are continuous at the boundary of
the medium. 
At $t=0$, the $E_1$ pulse is completely outside of the medium and
$E_2$ is zero everywhere. The numerical solution is easily obtained
using standard mathematical software packages such as Mathematica.
The results can be summarized as follows.

\begin{figure}[b!]
\begin{center}
\includegraphics{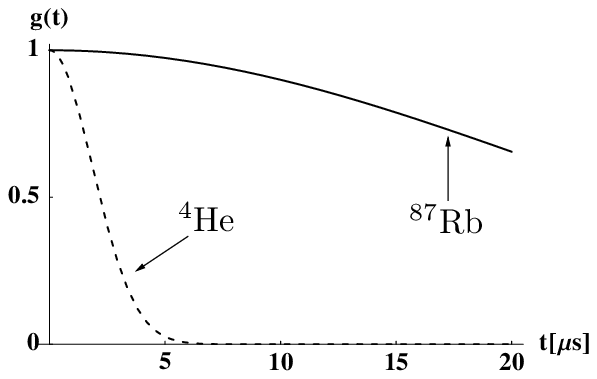}
\caption{\label{decay} \small Plot of  $g(t) = \exp[-2\,E_r^2\,t^2]$ as 
a function of time.  The recoil frequency of ${}^{87}$Rb 
($E_r = 2.3\times 10^4\,s^{-1} $) is about 10 times smaller than for 
 ${}^{4}$He ($E_r = 2.7\times10^5\,s^{-1}$).}
\end{center}
\end{figure}

The time dependence of $\varrho_0$ manifests itself solely in the 
presence of the function $g$. For a signal pulse whose length is 
short enough so that $g \approx 1$, the efficiency of the 
backscattering-effect is as high as for the idealized VSCPT-gas 
(Sec.~\ref{backscatter}). Fig.~\ref{decay}  
shows a plot of $g$ as a function of time for the Helium and 
Rubidium parameters. 

The small mass of ${}^{4}$He results in a 
large recoil frequency ($E_r = 2.7\times 10^5 s^{-1}$) and causes 
a rapid decay of $g$ within $5\,\mu s$. 
Consequently, the length of the incoming pulse must be as short as 
$2\times10^{-7}\,s$ in order to maximize the peak intensity of the 
backscattered pulse. However, one should bear in mind that the 
solution to the homogenous part of eq. (\ref{firstorder}) cannot 
be neglected for $ t\le 1/\gamma$. Therefore, the predictions of our 
theory are only correct for pulses of a few $\mu s$ or longer. 
In the case of Helium, the efficiency of the backscattering-effect 
is less than 1\% if a Gaussian pulse of $2\,\mu s$ is shone into the probe. 

A completely different behavior should be observed for a VSCPT gas of Rubidium atoms. 
The recoil frequency of Rubidium is about ten times smaller than for 
Helium, and $g$ is almost constant within the first $5\,\mu s$. 
Even for a Gaussian pulse of $4\,\mu s$, the peak intensity of the 
backscattered signal beam is almost as high as in the case of the 
idealized  VSCPT gas (section \ref{backscatter}). Thus, we expect that
for signal pulses traveling through a Rubidium gas the backscattered
beam should be a good signal for the entanglement between the 
electronic and center-of-mass degrees-of-freedom in a VSCPT state.
      
\section{Electromagnetically induced transparency \label{eit}}        
We  will now demonstrate that a VSCPT-gas creates
electromagnetically induced transparency 
\cite{hau1, hau2, fleischhauer1, fleischhauer2, scully,phillips1}
if the signal and the pump field interact with the gas at the same time. 
In presence of the pump field, the Hamiltonian in
Eq.~(\ref{masterequation}) has to be replaced by 
$H_{\text{C}} + H_\text{s}$. The same perturbative methods that 
have been employed in section \ref{backscatter} can be applied here, 
provided the signal field is much weaker than the pump field 
($ |\Omega_s| / |\Omega_p|\ll 1$). 
We consider the simplified situation where all atoms are 
initially in the unperturbed stationary state $|\text{VSCPT}\rangle$;
the previous section indicates that this is
justified for sufficiently short signal pulses.
The first order correction $\varrho_1$  is now determined by 
\begin{equation}
\big[\partial_t - \mathcal{L}_{\text{C}}\big]\varrho_1   = 
 -\frac{i}{\hbar}\left[H_\text{s},\varrho_0\right] \; ,
\label{firstorder2} \end{equation}
where $\mathcal{L}_{\text{C}}$ is defined as 
\[
\mathcal{L}_{\text{C}}(\,\cdot\,):=
-\frac{i}{\hbar}\,[H_{\text{C}},\,\cdot\,] \, + \, 
\mathcal{L}_{\gamma}(\,\cdot\,) 
\]
with 
$
H_{\text{C}} =  H_0 + H_p 
$. 
The unitary transformation  
\[
  U := \ket{e}\bra{e} e^{i\omega_p t}  + \ket{-}\bra{-} e^{i k_p \hat{z}} 
  + \ket{+}\bra{+} e^{-i k_p \hat{z}} 
\]
removes the operators $\exp[\pm i k_s \hat{z}]$  as well as 
the time dependence from the interaction Hamiltonian $H_p$. 
The transformed operator $\tilde{\varrho_1}:=U\varrho_1U^{\dagger}$ 
then obeys a more convenient equation than Eq.~(\ref{firstorder2}). 
The remaining calculation follows exactly the procedure of
Sec.~\ref{backscatter}. 
In particular, the matrix element $\bra{e}\varrho_1\ket{e}$ 
vanishes even in the presence of the pump field. 
This is a consequence of the initial atomic state 
$|\text{VSCPT}\rangle$ being an eigenstate
of the complete Hamiltonian $H_{\text{C}}$. 
If the finite width of the atomic momentum distribution is taken into account, 
excitations will occur and, consequently,  side bands of 
frequency $\omega_{\text{sb}} := \omega_s -2\,\Delta \omega$ will be present. 

The result for the field amplitudes is again of the form
(\ref{periodisch}) and (\ref{linear}) if one 
replaces $\delta$, $n_0$ and $\chi_0$ by 
$\delta_p := \sqrt{(2 n_p - \Delta k)\Delta k}$,
\begin{equation}
  n_p(\omega_s):= \frac{1}{2\,k_s}  
  \frac{\omega_s^2 \bar{\rho} |d|^2}{2 c^2 \varepsilon_0 \hbar}
   \chi_p(\omega_s)
\label{np}\end{equation}
and
\begin{equation}
  \chi_p(\omega_s) := 
  \frac{\Delta\omega}{ \Delta\omega\left( E_r +i\frac{\gamma}{2} + 
  \Delta_s \right) - 2 |\Omega_p|^2}\; .
\end{equation}
We exploited that 
$\hbar \Delta k^2/(2 m)$ and $\hbar k_p \Delta k/m $ 
are much smaller than 
the frequency difference $\Delta\omega:=\omega_s-\omega_p=c \Delta  k$ 
between the signal and the pump field. 

This result allows us to investigate the behavior of a 
$\sigma^+$ polarized signal pulse.  
The electrical fields $\mf{E}_1$ and $\mf{E}_2$ within the medium 
are then given by 
\begin{eqnarray}
\mf{E}_1^{(+)}(z,t)   & := &   
\int\limits^{\infty}_{-\infty} \mathcal{E}_1\,
e^{i\left(\frac{\omega}{c}\,z-\omega\,t\right)} \text{d}\omega 
\mf{\epsilon}^{(+)} \label{efeldpulse} \\[0.2cm]
\mathcal{E}_1(z, \omega )  & := & 
\frac{1}{\sqrt{2\pi}}\,\mathcal{E}_0(\omega) 
\left( 1 - \frac{n_p(\omega)\,z}{ n_p(\omega) a -i}\right) \nonumber
\end{eqnarray}
and 
\begin{eqnarray}
\mf{E}_2^{(+)}(z,t)  & := &  
\int\limits^{\infty}_{-\infty} \mathcal{E}_2 
e^{-i\left(\frac{\omega}{c} z + \omega t\right)} \text{d}\omega
\mf{\epsilon}^{(-)} 
 \label{efeldruck} \\[0.2cm]
\mathcal{E}_2(z, \omega ) & := & 
\frac{1}{\sqrt{2\pi}}\,\mathcal{E}_0(\omega)
\left(\frac{n_p(\omega)(a-z)}{n_p(\omega ) a - i}
\right) \quad. \nonumber
\end{eqnarray}
At $z=0$, just outside the medium,
the electric field $\mf{E}_1$ reduces to
\begin{equation}
\mf{E}_1(z=0,t) = E_0(t)\,\mf{\epsilon}^{(+)}\,+\,\text{c.c.} \quad,
\end{equation}
where  $E_0(t):= 1/\sqrt{2 \pi} \int_{-\infty}^{\infty}
\mathcal{E}_0(\omega) e^{-i\omega t}\text{d}\omega$ 
is the Fourier transform of  $\mathcal{E}_0(\omega)$.  
\begin{figure}[t!]
\begin{center}
\includegraphics{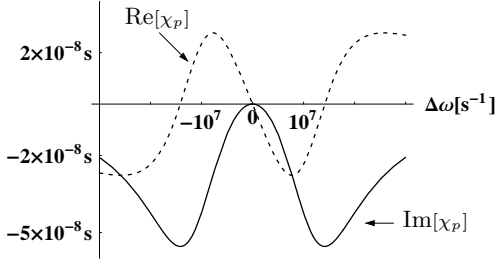}
\caption{ \label{dispersion} \small  Real and imaginary part of 
$\chi_p(\omega_s)$ for the following set of parameters:
$\Omega_p =  10^7\,s^{-1}$,
$E_r = 2.3\times10^{4}\,s^{-1}$ and 
$\gamma = 3.61\times 10^7\,s^{-1}$. 
We also assumed that the detuning of the pump field vanishes, 
i.e. $\omega_p=\omega_0$.}
\end{center}
\end{figure}

Fig.~\ref{dispersion} shows the real and imaginary part of $\chi_p(\omega)$ which is  
related to $n_p(\omega)$ through Eq.~(\ref{np}). 
Since $\chi_p(\omega)$ vanishes for $\Delta\omega=0$, 
the VSCPT-gas will be transparent for an incoming cw-field  of frequency 
$\omega_s=\omega_p$. The backscattered beam is then equal to zero, see 
Eqs.~(\ref{efeldpulse}) and (\ref{efeldruck}).  
An EIT-situation can thus be realized for an incoming signal pulse whose 
Fourier   components $\mathcal{E}_0$  
are sharply peaked around the pump laser frequency $\omega_p$.  
In order to evaluate the integrals in Eqs.~(\ref{efeldpulse}) and (\ref{efeldruck}) 
analytically, we assume that 
\[ \mathcal{E}_0(\omega):=\frac{E_0}{\sigma_{\omega}}\,
\exp\left[i\,4\frac{(\omega-\omega_p)}{\sigma_{\omega}}\right]\,
\exp\left[-\frac{(\omega-\omega_p)^2}{2\sigma_{\omega}^2}\right]\]
is given  by a Gaussian  
centered around $\omega=\omega_p$, the phase factor ensures that the $\mf{E}_1$ pulse 
is completely outside the medium at $t=0$. 
If the  width $\sigma_{\omega}$ of $\mathcal{E}_0$ is sufficiently small,  
$n_p(\omega)$ can be expanded as  
$n_p(\omega) \approx n_p'(\omega_p)\cdot\Delta \omega\;.$ 
With the help of the residue theorem we arrive at the following expressions 
for the field amplitudes 
\begin{eqnarray}
E_1(z,t)  & = & E_0\,\exp\left[-\frac{\kappa_+^2}{2\sigma_{\omega}^2}\right]\,
\left(1-\frac{z}{a} - h_+(\eta_+)\,\frac{z}{a^2}\right)\\[0.2cm]
E_2(z,t) & = & E_0\,\exp\left[-\frac{\kappa_-^2}{2\sigma_{\omega}^2}\right]
\left(1-\frac{z}{a} - h_-(\eta_-)\, \frac{z-a}{a^2}\right)\nonumber\;,
\end{eqnarray}
where
\begin{eqnarray}
h_{\pm}(\eta_{\pm}) &  =  & \sqrt{\frac{\pi}{2}}\,
\frac{e^{\eta_{\pm}^2}}{\sigma_{\omega} n_p'(\omega_p)}
\left(\text{Erf}(\eta_{\pm}) +1 \right) \\[0.1cm]
\eta_{\pm} & = & \left( \kappa_{\pm} + \frac{1}{a\, n'(\omega_p)}\right)\,
\frac{1}{\sqrt{2}\,\sigma_{\omega}} \nonumber \\[0.2cm]
\kappa_{\pm} & = & \left(t\, \mp \,\frac{z}{c} -\frac{4}{\sigma_{\omega}}\right)\,\sigma_{\omega}^2
\nonumber
\end{eqnarray}
and $\text{Erf}(\eta_{\pm})$ denotes the error function.

\begin{figure}[t!]
\begin{center}
\includegraphics{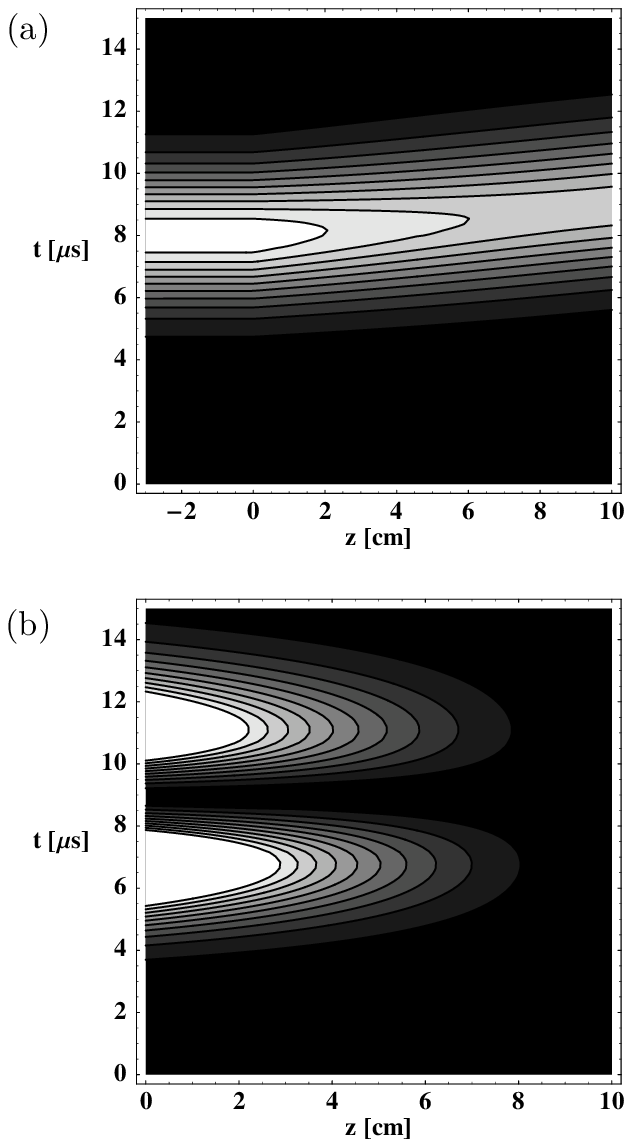}
\caption{ \label{pulsePic} \small Contour plots of the intensities of the 
incoming (a) and the reflected (b) pulse. 
The medium stretches from $z=0$ to $z=10\,$cm, the frequency width of the 
incoming pulse is given by 
$\sigma_{\omega}= 5\times 10^5\,\text{s}^{-1}$. We consider a 
gas of Rubidium atoms with a mean atomic density of 
$\bar{\rho}=2\times10^{10}\,\text{cm}^{-3}$, a Rabi frequency 
of $\Omega_p =  10^7\,s^{-1}$ and a zero detuning ($\omega_p=\omega_0$) 
for the pump lasers. The trajectory of the incoming pulse in (a) shows a bend at $z=0$.  
This is  a consequence of the reduced group velocity inside the medium.}
\end{center}
\end{figure}

Fig.~\ref{pulsePic} (a) shows a contour plot of the intensity of the  $\mf{E}_1$ pulse 
just before and inside the medium, which stretches   from $z=0$ to $z=10\,$cm. 
This unrealistic probe length has been chosen to better visualize  
the  reduced group velocity of the $\mf{E}_1$ pulse within the medium, 
which also appears in other 
EIT-media \cite{hau1,phillips1,fleischhauer1}.
Starting from $z=0$, the trajectory of the incoming pulse is tilted torwards the positive $t$-axis 
which is a consequence of the group velocity reduction inside the VSCPT-gas. 
Fig.~\ref{pulsePic} (b) shows the intensity of 
the backscattered beam $\mf{E}_2$.  It can be seen that the incoming pulse gives rise to 
two backscattered beams which can be understood as follows.  Since $n_p(\omega)$ vanishes for $\omega=\omega_p$,  
the Fourier components close to $\Delta\omega=0$ do not contribute in the integral in Eq.~(\ref{efeldruck}). Consequently, 
the Gaussian $\mathcal{E}_0$ is split into two parts which build up the two reflected beams.  
We finally note that an increased atomic density $\bar{\rho}$, a broadened frequency 
width $\sigma_{\omega}$ or a less intense pump field  amplifies the backscattering effect and 
reduces the transparency for the incoming pulse. For $\bar{\rho}=10^{11}\,\text{cm}^{-3}$, $\sigma_{\omega}=10^6$, 
$\Omega_p=5\times 10^{6}$ and a probe length of $a=1\,$cm,   the peak intensity of the reflected beam  
(relative to the peak intensity of the incoming pulse) 
is about 0.3, for the parameters of Fig.~\ref{pulsePic} it is given by 0.08. 

\section{Quantization of the signal field}\label{secqft}
To see how a VSCPT gas influences the quantum state of the signal beams
we consider the evolution of the field operator for the two signal modes 
\cite{interaction},
\begin{equation}
  \hat{\mf{D}}_{\text{s}} :=    
\varepsilon_0  \,i \mathcal{E}_{\omega_s} \sum\limits_{i=1}^2 a_i(t)
  \mf{\epsilon}_i e^{i k_i z} + \mbox{H.c.} \; ,
\label{fosys}
\end{equation}
where the modes are characterized by $k_1 = k_s$, $k_2=-k_s$ and
$\mf{\epsilon}_1 = \mf{\epsilon}^{(+)} $ as well as
$\mf{\epsilon}_2 = \mf{\epsilon}^{(-)} $.
We have set $\mathcal{E}_{\omega_s}:=\sqrt{\hbar\omega_s/(2\varepsilon_0 V)}$ 
with $V$ being the quantization volume.
We seek a solution to Heisenberg's equation of motion 
for the (transverse) displacement $\hat{\mf{D}}_{\text{s}}$, 
which takes on the form 
\begin{equation}
  \left[\frac{1}{c^2} \partial_t^2-\Delta\right]\hat{\mf{D}}_{\text{s}} 
  = 
  \nabla \times \nabla \times \hat{\mf{P}}_{\text{s}}\; ,
\label{heisenberg}
\end{equation} 
where  $\hat{\mf{P}}_{\text{s}}:= 
\mbox{Tr}_{\text{R}}(\varrho_{\text{R}}\hat{\mf{P}})$ 
as well as $a_i(t)$
are reduced Heisenberg operators \cite{decoherence}.
The trace runs over all atomic and radiation
degrees-of-freedom except for the two signal modes. 
It turns out that the operator $\hat{\mf{P}}_{\text{s}}$ has the same
form as the classical polarization (\ref{fullp}) if one replaces 
the Rabi frequencies $\Omega_{1,2}$ by 
$i d \mathcal{E}_{\omega_s} a_{1,2}/\hbar$. 
Introducing the slowly varying operators    
$ \tilde{a}_i(t) := a_i(t) \exp (i \omega_s t)$ one can derive
the coupled set of first-order differential equations 
\begin{equation}
  \partial_t \left( \begin{array}{c} \tilde{a}_1 \\ \tilde{a}_2
  \end{array} \right) 
  = i\beta \left( \begin{array}{rr} -1 & 1 \\ 1 & -1 \end{array} \right )
  \left( \begin{array}{c} \tilde{a}_1 \\ \tilde{a}_2 \end{array} \right )
  \; , \label{sys3}
\end{equation}
where second derivatives of the operators $\tilde{a}_{1,2}$ have been 
neglected against $\omega_s\dot{\tilde{a}}_{1,2}$. 
The coefficient $\beta$ is given by 
$\beta  := c \, n_x$, where $n_{x}$ can  be either  
$n_0$ of Eq.~(\ref{nnull}) or $n_p$ of Eq.~(\ref{np}).
We arrive at the following expression for the reduced annihilation operators
\begin{eqnarray}
a_1(t) & = & e^{-i(\beta + \omega_s) t}\,\left[\cos(\beta t)\,a_1
+  i \sin(\beta t)\,a_2 \right] \label{aheisenberg} \\[0.1cm]
a_2(t) & = & e^{-i(\beta + \omega_s) t}\,\left[i \sin(\beta t)\,a_1
+   \cos(\beta t)\,a_2 \right]\; , \nonumber
\end{eqnarray}
where the operators $a_i$ on the right hand side are annihilation operators in the 
Schr\"odinger picture.
The operators $a_i(t)$ do not obey the canonical commutation relations
since $\beta$ is a complex parameter.
This result is consistent,  since canonical commutation
relations are not required for reduced Heisenberg operators
\cite{decoherence}. 

Solution (\ref{aheisenberg}) demonstrates that a single photon in the
incoming signal beam $a_1$ will evolve into a superposition of
the two signal modes. Such a state corresponds to 
single-particle entanglement, 
since the polarization and position degrees-of-freedom of the photon
are then entangled. Hence, the one-particle entanglement which is
present in the atomic VSCPT state can be transferred to a corresponding
entanglement of a photon in a weak signal pulse.
We remark that, apart from the possible appearance of EIT,
the creation of single-particle entanglement could also be achieved
by a beam splitter followed by a polarization rotator in one of 
the two output modes. The distinguishing feature of the VSCPT gas is that
this effect is a direct signal of atomic entanglement.

In conclusion, we have shown that an atomic gas prepared in a VSCPT
state exhibits unique optical features which include the phenomenon
of EIT and a backscattered light pulse which is a signal for the 
entanglement associated with the VSCPT state. Detection of the backscattered
light pulse should be possible for a gas of Rubidium atoms, while
the large recoil velocity in Helium would make such an experiment
unfeasible. On the few-photon level, the VSCPT gas would
lead to single-particle entanglement for a signal photon.

\textbf{Acknowledgments}\quad
We would like to thank B.~C.~Sanders for fruitful discussions. 
This work was supported by the German Academic Exchange Service (DAAD) and 
Alberta's informatics Circle of Research Excellence (iCORE).



\begin{thebibliography}{99}
\bibitem{remark1}{To a first approximation the population of the excited and the bright 
state (the one that is orthogonal to $\ket{D(p)}$) can be neglected.}
\bibitem{linearity}{Since the equation for $\varrho_1$ is an inhomogeneous differential
equation with an inhomogeneity linear in $E_1$ and $E_2$,
one can simply  treat the two cases separately and add the two
inhomogeneous solutions afterwards.}
\bibitem{neglect}{The paraxial approximation for $\mf{E}_2$ takes on a different 
form than for $\mf{E}_1$ since $k_2^2 c^2\not =\omega_s^2$. 
Terms of the order of magnitude $\Delta k^2/k_s$ have been 
neglected.}
\bibitem{boundary}{Maxwell's equations impose boundary conditions on the 
electric and magnetic fields $\mf{E}$ and $\mf{B}$ at $z=0$ and $z=a$ which  
result in a reflection of the fields at these boundaries.  
In the paraxial approximation, this reflection is negligible and the boundary  
conditions for the electric field reduce to Eq.~(\ref{boundary}).}
\bibitem{vgroup}{The time dependence of 
the amplitudes  $E_{1,2}$ has not 
been taken into account in the calculations for the denity operator. 
This amounts to ommit the changed  group velocity of the pulse within the gas.}
\end{thebibliography}
\end{document}